# Spacecraft Tracking Applications of the Square Kilometre Array

J.G. Bij de Vaate[(1)], L.I. Gurvits[(2)], S.V. Pogrebenko[(2)], C.G.M. van 't Klooster[(3)]

| | | |
|---|---|---|
| [(1)]*ASTRON* | [(2)]*JIVE* | [(3)]*ESA-ESTEC* |
| *Oude Hoogeveensedijk 4* | *Oude Hoogeveensedijk 4* | *P.O. Box 299* |
| *7991 PD Dwingeloo* | *7991 PD Dwingeloo* | *2200 AG Noordwijk* |
| *The Netherlands* | *The Netherlands* | *The Netherlands* |
| *Email: vaate@astron.nl* | *Email: lgurvits@jive.eu* | *Email: Kees.van.t.Klooster@esa.int* |

## ABSTRACT

The Square Kilometre Array (SKA) is the next generation radio telescope distinguished by a superb sensitivity due to its large aperture (about one square kilometre) and advanced instrumentation. It will cover a broad range of observing bands including those used for tracking of and communications to deep space missions. While spacecraft tracking is not a main application defining the technical specifications of the SKA, this facility might play a role in tracking deep space probes as a backup to the "dedicated" deep space tracking networks. This paper presents possible applications of the SKA as a deep space tracking facility and major related technical specifications of various concepts of the SKA. It was presented at the 3$^{rd}$ International Workshop on Tracking, Telemetry and Command Systems for Space Applications, ESA–ESOC, Darmstadt, Germany, 7–9 September 2004. Over the past years, the SKA concept has developed to a much higher level of detalisation and is currently at the implementation phase. A number of specific considerations in this presentation no longer correspond to the actual status of the SKA project. However, the overall concept of the SKA applications for communication and tracking of interplanetary spacecraft remain topical, and some approaches presented here remain of interest for prospective deep space missions.

## 1  INTRODUCTION

Radio astronomy instrumentation currently rely on the use of large paraboloidal reflector antennas often larger then 25 m diameter and at some locations setup in arrays in order to increase the baseline up to several to tens of kilometres. Very Long Baseline Interferometry (VLBI), where antennas are combined on the scale of several hundred to thousands km or even with satellite based antennas, increases the angular resolution to sub-milliarcseconds. Limitations arise because these existing telescopes typically have collecting areas below 10.000 m$^2$ while observing with a single beam and limited instantaneous frequency bandwidth. An initiative has emerged to develop a telescope to provide two orders of magnitude increase in sensitivity over existing facilities at metre to centimetre wavelengths. To achieve this goal it will require a telescope with one square kilometre of collecting area. Extensive discussion of the science drivers and of the evolving technical possibilities has led to a concept for the Square Kilometre Array and a set of design goals. SKA will be an interferometric array of individual antenna stations, synthesizing an aperture with diameter of up to several thousand km. A number of configurations are under consideration to distribute the 10$^6$ square meters of collecting area. These include 30 stations each with the collecting area equivalent to a 200 metre diameter telescope, and 150 stations each with the collecting area of a 90 m telescope. Approximately 50% of the array is to be contained within a centrally-condensed inner array to provide ultrahigh brightness sensitivity at arc-second scale resolution. The outrigger stations provide a ten to one hundred-fold increase in angular resolution to enable high-resolution imaging [1].

## 2  SQUARE KILOMETRE ARRAY SCIENCE REQUIREMENTS

The international SKA community generated a set of so called 'level 0', or highest priority, science goals for the SKA. These are based on 'traditional' astronomy science projects but also planetary science and SETI. From these level 0 science projects, SKA system requirements have been generated. Table 1. [2] gives a short-list focussed on the requirements that are relevant for the Deep Space Network (DSN). Obvious, DSN is not a level 0 science (also called Key Science Projects). However the higher frequency limit of 35GHz, listed as goal is driven by potential DSN application *and* a science case. Important in this table is that none of the Key Science Projects requires multiple fields of view (FoV). The goal of 4 independent FoV would enhance observing flexibility and speed of all large-scale surveys. Some observing modes such as pulsar timing and wide angle astronomy require multiple FoV, but do not require the full SKA sensitivity. For these cases sub-arraying is sufficient. The instantaneous pencil beams are beams that are 'locked' within the primary field of view, e.g. 1 square degree at 1.4 GHz.

## Table 1. SKA science requirements

| Parameter | Requirement |
| --- | --- |
| Total Frequency Range | 100 MHz – 25 GHz, goal: 60 MHz – 35 GHz |
| $A_{eff}/T_{sys}$ (at 45 degrees elevation) | $2 \times 10^4$ m$^2$/K between 0.5 and 5 GHz<br>$1.5 \times 10^4$ m$^2$/K between at 15 GHz<br>$1.0 \times 10^4$ m$^2$/K between at 25 GHz<br>$0.5 \times 10^4$ m$^2$/K between at 35 GHz (goal) |
| Simultaneous independent observing bands | 2 pairs with 2 polarizations |
| Max freq. Separation of observing bands | Factor of 3 |
| Number of separated field of view (FoV) | 1 with full sensitivity        goal: 4 with full sensitivity<br>10 simultaneous sub-arrays |
| Configuration | 20% of total collecting area within 1 km diameter, 50% of total collecting area within 5 km diameter, 75% of total collecting area within 150 km diameter, maximum baseline at least 3000 km |
| Imaging Field of View | 1 square deg. @ 1.4 GHz |
| Number of Instantaneous Pencil Beams | 50 (within one FoV) |
| Angular Resolution | 0.02 arcsec @ 1.4 GHz |
| Instantaneous Bandwidth | 25% but not more then 4 GHz |
| Imaging Dynamic Range | $10^6$ @ 1.4 GHz |
| Polarization Purity | –40 dB |

## 3     SQUARE KILOMETRE ARRAY CONCEPTS

The international community is working on the realization of the SKA since the conception of the idea in 1997. Triggered by first discussions held in Europe, e.g. at workshops as an ESA Antenna Workshop on Large Antennas for Radio Astronomy, 1996. A total of 7 concepts have been created that all can fulfil a good part of the SKA system requirements. The different institutes have build demonstrators in order to prove, or at least analyse some aspects, of these concepts. The concepts can roughly be divided in two categories, concepts with large antennas and concepts with small antennas. All concepts are based on the assumption that the traditional 25 meter arrays, like the Very Large Array and the Westerbork Synthesis Array, cannot cost effectively be extended to a SKA.

None of the system concepts is compliant with all the SKA system requirements. The final SKA design will therefore most likely be a hybrid design; a combination of two or more concepts. The International SKA Steering Committee (ISSC) requested the (national) institutes to demonstrate the feasibility of their concept with a 1% SKA demonstrator. The results of the SKA demonstrators, to be operational by the end of 2007, will lead the creation of a short list with preferred concepts and hybrid solutions.

*The Large-N small-D Concepts*

The most straightforward way of building a large collecting area is known as the Large-N small-D concept (LSDN), pursued by the US consortium. A large number, over 4000, of 12 meter dishes are being proposed. A smart feed design should make a large bandwidth possible, 0.15 to 34 GHz. The small diameter approach is seen as more cost effective then 'just' building a large number of 30-50 meter telescopes.

A deviation of the US approach is the 12 m dish proposed by the Radio Astrophysics Institute of India. In the contrary to the US solid dish, a wire mesh preloaded parabolic dish (PPD) will be designed. With the advantage of lower fabrication cost and, due to the lower wind load, engine control complexity. The maximum frequency is however limited to 10 GHz.

*The Large Adaptive Reflector*
The Large Adaptive Reflector (LAR) concept is being developed by the National Research Council of Canada. The LAR will consist of an array of 60 200-m Reflector Antennas where the methodology behind the LAR idea is to:
- Utilize a reflector antenna, with all its favourable properties for frequency-independent radio wave reflection
- Build the reflector with a large focal ratio (f/d, where f is the focal length) so that the curvature of the reflector is very small
- Provide an independently moveable focal platform and an adjustable reflector shape to provide "access" to a whole family of Zenith Angle. The goal is to cover a Zenth Angle range up to 60 at all Azimuths. The platform will be elevated with an aerostat attached to cables with wenches for position adjustment.

Expected frequency range of the LAR system is 0.15 to 22 GHz

*The Luneburg Lens*
The Luneburg lens, a spherical radio lens as the first stage beamformer, is being studied by CSIRO, Australia. The Luneburg approach, with its wideband optical beamforming and intrinsic capability for placing multiple beams across the sky, is an intermediate one offering some of the signal processing flexibility associated with phased arrays, as well as most of the performance and versatility of reflecting concentrators. Multiple beams can be placed on the sky independently. The frequency range is limited by the loss of the dielectric, on the high end, and by the size of the lens, on the low end. Estimated is currently 0.1 to 10 GHz. The SKA should contain 19.000 7 m lenses, which will create a square kilometre collecting area only if you take the multi-beaming capability into account.

*The Cylindrical Reflector*
The Cylindrical Reflector is the second concept, which is being studied by CSIRO, Australia. Large cylindrical reflectors are advantages compared to phased array's in terms of cost; less receivers and advantage compared to paraboloids in terms of field of view; more beams are possible on a larger area of the sky. The multiple beams, to be synthesized with a smart (digital) line feed, are however linked: the beams are not completely independent.
Frequency range is believed to be 100 MHz to 9 GHz, where higher frequencies are possible (up to 20 GHz) with sensitivity loss.

*KARST*
The National Astronomical Observatories of China (NAOC) is proposing a small number of large reflecting surfaces installed in a naturally occurring depression, with the receiver system suspended from cables on pylons at the edge of the depression. The concept is called the KARST (Kilometer square Aperture Radio Synthesis Telescope). The KARST consists of 30 300 m diameter reflectors, with a frequency range of 100 MHz to 2 GHz (extension to 8 GHz might possible).

*Phased Arrays*
The European SKA Consortium is pursuing the concept of aperture arrays. Each station in this concept consists of an array of integrated antenna tiles. Each tile consist of a collection of simple, all sky antenna elements coupled such that beams are formed and steered electronically on the sky: the phased array principle. Although this concept is the most advanced of those under investigation, besides unique features, it has serious drawbacks. An antenna element spacing significantly above $\lambda/2$ should be avoided since grating lobs will degrade the efficiency of the array. Operating frequencies above 2 GHz are therefore considered not to be practical due to the extreme number of antenna elements required. The electronic steering and beamforming creates the possibility of generating multiple independent beams. Which makes the aperture array a multi-user instrument, a very important aspect for an instrument of this size and cost. The effective observing time can be expended without the loss of sensitivity.
Fig. 1 gives the system characteristics for a two-array phased arrays, both with an approximate bandwidth of 3.5:1. The second array does not meet the 20.000m$^2$/K, it is expected not to be cost effective to build a full SKA for the higher part of the band. This two-band approach maybe adopted by the European SKA Consortium, however the straw man design is a three-band concept.

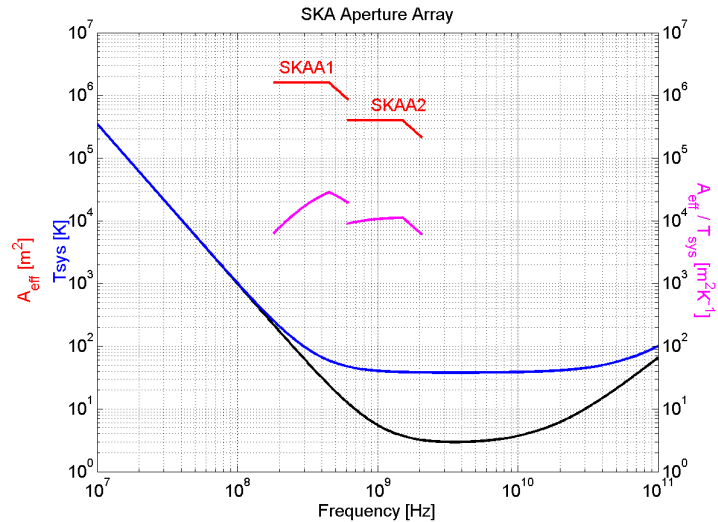

*Fig. 1.* System characteristics of a two-array phased array concept proposal

Table 2 gives an overview of the capabilities of the 7 described concepts with respect to the crucial criteria for Deep Space Tracking and Navigation. These are not the only crucial criteria, but these where the differences between the station concepts are dominant.

The frequency of 2.4 GHz is listed, although it is planned to be phased-out, for completeness but also since the multi-beam capabilities of the phased-array concept in this band might generate new potential for it. The columns $A_{eff}$ give an indication of the effective area at the given frequency. A reduction will influence the sensitivity of the system but not aspects as angular resolution. Since the Phased Array concept will most likely be designed as a dense array for frequencies up to 1.4GHz, at 2.4GHz a reduction in the $A_{eff}$ of this concept should be anticipated. Except for the Phased Array concept most approaches will be capable or can be designed for 8GHz reception as well as 2.4GHz. However due loss in efficiency, a lower $A_{eff}$ is expected for some concepts at 8GHz as well. Only the small dish concept (Large-N Small-D) foresees a good functionality at 32GHz.

The column *multi-beams* gives the capabilities of the concepts for creating Multiple Fields of View (FoV). Only the Phased Array and the Luneburg lens concept are capable of Multiple FoV's, while maintaining the full sensitivity of the complete system for each FoV. All concepts do support multibeaming within a primary beam, e.g. 1º square at 1.4GHz, which could be an advantage for tracking multiple spacecrafts around a planet.

**Table 2.** Cross-reference comparison matrix

|  | 2.4GHz | $A_{eff}$ @ 2.4 GHz | 8 GHz | $A_{eff}$ @ 8 GHz | 32 GHz | Multi-beams (fov) | Sky coverage |
|---|---|---|---|---|---|---|---|
| Large-N Small-D | Yes | 100% | yes | 100% | yes | no | good |
| Phased Array | Maybe | 20% | no | - | no | yes | good |
| LAR | Yes | 100% | yes | 90% | no | no | limited |
| Luneburg lens | Yes | 50% | maybe | 10% | no | yes | good |
| Cylindrical array | Yes | 100% | yes | 90% | no | no | good |
| KARST | Yes | 100% | maybe | - | no | no | limited |
| PPD dishes | Yes | 100% | Yes | 25% | no | no | good |

A solution for the other concepts is the use of sub-arraying. Each sub-array forms a single FoV. The created beams have a reduced $A_{eff}$, however might still have the resolution of the complete system. Sub-arraying is very good possible with the large number of small reflectors, the very large reflectors, LAR and KARST, cannot be used very efficiently for sub-arraying due to small number of antenna elements.

The *sky coverage* of the large reflectors is limited due to mechanical constraints of these concepts. A limited sky coverage can be solved for astronomy with a dedicated scheduling, however is likely to cause a problem for Deep Space Telemetry.

## 4    DEEP SPACE TRACKING AND TELEMETRY WITH SKA

Although a Deep Space Network (DSN) is not seen as a level-0 science criteria, the international SKA community has set-up a working group to investigate the possibilities of Space Craft Tracking with SKA [3]. A SKA size array would allow of the order of 100 times greater data rate to the outer planets, smaller and less expensive spacecraft, longer missions in the case of Mars (where the distance varies from 0.33 to 2.5 AU), and very accurate real-time navigational data.

*Angular resolution:*
The maximum baseline length of at least 1000 km for angular tracking will be met by the SKA and will result in an angular resolution of 0.5 milliarcseconds (3000 km baseline) at 32 GHz. The two lower communication frequencies will have the respectively lower resolution.

*Continues data transmission:*
For telemetry reception from deep space missions continuity is critical, which will be an additional requirement for the back-end of SKA since for radio astronomical signals gaps can be allowed. And will appear due to phase-switching and block processing. It would be advantageous, if future missions where designed such that this requirement would not be applicable. Anyway, continuous 24 hours coverage will require 2 or 3 SKAs.

*Frequency range:*
As can be seen in table 3 most concepts can receive the 2.4 and 8 GHz signals. However only the Large-N Small-D concepts is capable of receiving 32 GHz signals.

**Table 3.** Summary specifications for DSN reception. Except for the first three items in table 3 the DSN reception generates very specific design requirements that should be take into account early in the design process and influences the cost of SKA.

| Item | Specification |
|---|---|
| Frequency coverage | 2.2–2.3 GHz, 8.4–8.5GHz, 31.8–32.3 GHz |
| Minimum sensitivity | 5000 m$^2$/K |
| Polarization | Right hand Circular (RCP) and Left hand Circular (LCP) |
| Data transmission | No gaps allowed |
| Baseline | >1000km |
| Reception | Simultaneous 2 and 8 GHz  Or Simultaneous 8 and 32 GHz (for reception calibration) |
| Timing | High time resolution |
| Configuration | Availability of "analogue sum" output from the central part of the array for each of the several beams |

*Sensitivity:*
The bit rate of the downlink telemetry signal is directly proportional to the G/T. Although even the Small-D Large-N concept will most likely only achieve an $A_{eff}/T_{sys}$ of 5000 m$^2$/K at 32 GHz, this will still give an G/T more then 10 dB higher compared to an G/T for $A_{eff}/T_{sys}$ of 20.000 m$^2$/K at 2.3 GHz. These sensitivities are valid if the complete square kilometre array can be used for DSN. This is only possible on more then an exceptional case if a SKA system has multiple independent beams for the frequency of communication. If independent multiple beams are not available only particular events can make use of the complete SKA. Higher Sensitivity creates a new view on the telemetry, lower power transmission requirements of the spacecraft allows for smaller design, lower power consumption and longer range communication.

**4.1 Possible Solution for DSN**

The most suitable candidate SKA configurations, emerging from present architectural studies, suggest a hybrid assembly consisting of various types of receiving elements, to provide sufficient overall contiguous frequency bandwidth operation. This is based on the use, for example, of phased array tiles for the lower frequencies and relatively small reflector dishes for the higher frequencies. In this context, it is especially the 12 m class dishes that could support telemetry and tracking in the S, X and Ka-band frequency bands.

Antennas with Gregorian feed systems will employ ultra-wide-band log-periodic antennas that could support simultaneous reception of at least two telemetry bands. Note that a coaxial three-band feed might be more applicable for a system designed for DSN only. In a hybrid configuration, it realistically planned to have 2000 to 4000 dishes, of which about a quarter are expected to be located within a central core of 2km diameter. The other three quarters of the dishes will be distributed over a few hundred clusters up to 1000 km away from the central core. For telemetry and navigation applications there will be at least 500 12 m class dishes in the central core available that can be used to provide an order of magnitude improvement in sensitivity over a 70m dish antenna. Within the SKA project, the cost for 500 of these 12 m dishes, together with the receivers, land acquisition, transport network and central processing node costs are estimated to US$ 150 millions. There will be additional cost to make the antenna combining system suitable for streaming telemetry data. The signals of the elements will be combined in a beamforming network that will have typically 800 MHz bandwidth, which is enough to support the full telemetry bands. For tracking, dynamic phase and time delay adjustment between antenna and summing point of the beam-former, will be required.

**5      VLBI TRACKING OF THE HUYGENS PROBE: AN EXAMPLE OF TRACKING AND RADIO ASTRONOMY SYNERGIES**

An assessment study of VLBI (Very Long Baseline Interferometry) observations of the Huygens probe during the probe's descent through the atmosphere to the surface of Titan [4] has been performed. The aim of the study was to assess the feasibility of a direct receipt, detection and VLBI processing of the probe's S-band radio signal. The direct receipt of the probe signal by Earth-based tracking stations was not foreseen in the original mission scenario but has proven to be possible owing to recent developments in radio astronomy, and particularly in VLBI. We analysed the power budget of the "Huygens–Earth" radio link, the potential accuracy of the VLBI determination of the probe's coordinates in the atmosphere of Titan, and some scientific applications of these measurements.

VLBI tracking of the Huygens probe in the atmosphere of Titan is based on the possibility to determine the angular position of a source of radio emission with an accuracy of the order of

$$\Delta\varphi \cong \frac{1}{SNR}\frac{\lambda}{B} \; , \tag{1}$$

where $\lambda$ is the signal wavelength, $B$ is the radio interferometer baseline, and *SNR* is the signal-to-noise ratio of the detected interferometric response. At the S-band frequency of 2 GHz and global baselines of the order of the Earth's diameter, the achievable accuracy is at the milli-arcsecond (mas) level. With SNR ~20–30 or more this corresponds to a sub-km accuracy of the probe position at the distance to Titan at the time of Huygens-Titan encounter, ~8 astronomical units (AU). Recent developments in the techniques of VLBI permit us to achieve this level of accuracy.

To derive the probe's position from the detected VLBI responses, the so-called phase-referencing technique [5] must be applied. This technique is based on the determination of the telescope phase offsets and propagation phase distortions by observing known bright references sources, deriving the phase corrections from these, and applying them to the data of a weaker target source. The achievable accuracy of phase calibration depends on the signal-to-noise ratio of the interferometer response to the calibrator source and its angular distance from the target. The most efficient use of phase referencing would be realised if both the target and reference source(s) were within the primary beams of all the participating radio telescopes. Nevertheless, it has been proven [6,7] that this switching technique works at angular separations between reference and target sources of up to several degrees, although it is less accurate as compared to 'in-beam' technique, depends on atmospheric conditions, and requires 'nodding' of telescopes between target and reference sources, which decreases the observing efficiency.

For the purpose of our experiment we are focusing on detection of the probe's narrow band signal carrier wave, because it represents the signal component with the maximum spectral power density and can provide the best possible signal-to-noise ratio. On the other hand, observations of calibrators require as wide a bandwidth as possible to achieve the desired accuracy. Digital disk-based VLBI recorders such as Mk5 will allow us to record the signal from both the probe and the calibrator onto the same medium, to ensure the coherency of phases between these two signals.

Observations of the probe's descent will be performed using VLBI telescopes capable of receiving the probe's signal at 2.040 GHz (with original receivers or specially tuned ones), although the telescopes equipped with standard S-band VLBI receivers (typically covering the frequency band above 2.2 GHz) will be used as well, in order to improve the phase calibration accuracy.

To process the data, we will need both wide bandwidth processing facilities for the calibrators' signals and ultra narrow band processing tools for the probe's signal. The EVN Mk4 VLBI correlator at JIVE [8] can process the broad band data from as many as 16 VLBI stations simultaneously, and can be enhanced with ultra-high spectral resolution hardware/software tools for narrow-band processing.

**5.1 Probe's Signal Detect Ability and Estimate of the Tracking SNR**

The effective isotropic power of the probe's Channel A transmitter is 10.2 – 11.0 W, and the transmitting antenna gain is 3.44 dBi (worst case scenario) in a 120 degrees cone [7]. The modulation scheme PM/BPSK/PCM-NRZM with 1.34 radians modulation index produces an output of ~7 W in a data band, and leaves ~3.5 W in a carrier wave. This yields a carrier wave signal power density (flux) at Earth (distance 8.2 AU) of $P_s$=4.8 $10^{-25}$ W/m$^2$.

The typical spectrum of a BPSK signal is shown in Fig.2. The data band with a 16 Kbps capacity is separated from the carrier by 130 KHz, which makes the carrier wave very well isolated from the data band and suitable for a high resolution radio spectroscopic analysis.

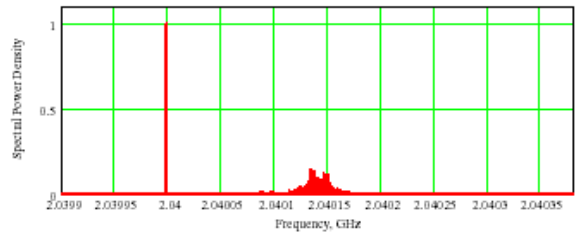

*Fig.2.* Typical spectrum of the BPSK modulated signal.

**Table 4.** Expected detection SNRs for typical stations at different integration times

| Station | D (m) | $A_{eff}$ | $T_{sys}$ (K) | SNR 1s | SNR 10s | SNR 100s |
|---|---|---|---|---|---|---|
| GBT | 100 | 0.71 | 23 | 7.70 | 77.0 | 770.0 |
| Parkes | 64 | 0.55 | 35 | 1.50 | 15.0 | 150.0 |
| Usuda | 64 | 0.55 | 40 | 1.45 | 14.5 | 145.0 |
| VLBA_OV | 25 | 0.55 | 30 | 0.29 | 2.85 | 28.5 |
| VLBA_SC | 25 | 0.48 | 40 | 0.19 | 1.95 | 19.5 |
| Kashima | 34 | 0.65 | 75 | 0.25 | 2.50 | 25.0 |
| Mopra | 22 | 0.60 | 37 | 0.19 | 1.96 | 19.6 |
| Kashima_11 | 11 | 0.80 | 72 | 0.03 | 0.33 | 3.30 |

High resolution spectral analysis combined with proper phase correction of the received signal can yield a detection *SNR* at a single station with antenna diameter *D*, aperture efficiency $A_{eff}$, system temperature $T_{sys}$, coherent integration time $t_{int}$ and frequency resolution $\delta F$, ($\delta F = 1 / t_{int}$):

$$SNR = A_{eff} \frac{\pi D^2}{4} \frac{P_s}{k T_{sys}} \frac{1}{\delta F} \quad , \tag{2}$$

where $P_s$ is the carrier wave flux at Earth and $k$ is Boltzmann's constant. Expected SNR values for different integration times for a set of typical telescopes which can *in principle* observe the Huygens probe are listed in Table 4.

A signal-to-noise ratio $SNR_{XY}$ on a baseline between telescopes $X$ and $Y$ in the case of a point source and perfect phase calibration is

$$SNR_{XY} = \sqrt{SNR_X \, SNR_Y} \quad , \tag{3}$$

where $SNR_X$ and $SNR_Y$ are those for stations $X$ and $Y$. The results for cross-correlation SNRs for 8 typical stations (the same as listed in Table 4) and 50 s coherent integration time ($\delta F$=20 mHz) are listed in Table 5.

**Table 5.** Baseline SNR's for 50 s coherent integration on various baselines

| Stations | 1 | 2 | 3 | 4 | 5 | 6 | 7 |
|---|---|---|---|---|---|---|---|
| 1 GBT | | | | | | | |
| 2 Parkes | 170 | | | | | | |
| 3 Usuda | 167 | 74 | | | | | |
| 4 VLBA_OV | 74 | 33 | 32 | | | | |
| 5 VLBA_SC | 61 | 27 | 26 | 12 | | | |
| 6 Kashima | 69 | 31 | 30 | 13 | 11 | | |
| 7 Mopra | 61 | 27 | 27 | 12 | 10 | 11 | |
| 8 Kashima_11 | 25 | 11 | 11 | 5 | 4 | 5 | 4 |

Estimated baseline SNR's are acceptable, even for small 11 m antennas, when correlated against 100 or 64 metre class antenna. This leaves us with the key issue of achieving an adequate phase calibration accuracy.

**5.2 Global VLBI Stations Suitable for Observations of the Probe**

At 09h ET on the 14<sup>th</sup> of January 2005, the Huygens probe is scheduled to start its transmission. About 70 minutes later the signal will reach the Earth. Fig.2 illustrates how the Earth will be visible from Titan during the probe's descent in the atmosphere.

During the probe's descent, Titan will be visible by Australian, Chines, Japanese and US radio telescopes. European IVS station Ny Ålesund will also see Titan. It is essential that at the beginning of the observations the probe will be visible by two large telescopes, GBT and Usuda. After an hour and a half Titan will go below the horizon for the GBT and other Eastern USA telescopes but will appear above the horizon for Australian telescopes such as Parkes, Mopra, Hobart and EVN telescopes at Shanghai (Sheshan) and Urumqi (Nanshan) in China. At the end of the expected probe's battery life the signal could be received by European telescopes such as Effelsberg, Ny Ålesund and Onsala. Potentially, the longest possible global baselines can be achieved during the whole duration of the mission. About 30 VLBI stations around the world could participate in the Huygens observations, provided they are equipped with 2.04 GHz receivers and disk based data recorders. Of course, the actual observing network is subject to special considerations.

### 5.3 VLBI Contribution to the Huygens Mission Scientific Return and the Prospects for Future Experiments

VLBI tracking of the Huygens probe during its descent through the atmosphere of Titan, would yield independent direct detection of the probe's signal on Earth and 3D determination of the probe's position in the ICRF frame every 10–30 seconds. These data will directly impact the interpretation of all other measurements on-board the probe. They also provide a useful synergy to the Doppler measurements of the Huygens motion both from the Cassini spacecraft and ground based stations.

As it is clear from the study described above, VLBI tracking of watt-level transponders at the distances of the order of 10 AU is a challenging but realistic task for present day VLBI technology. The critical parameter determining feasibility of such observations is sensitivity of the VLBI array. In the coming decades this parameter is likely to improve dramatically, due to the implementation of the next generation radio telescope the SKA. VLBI tracking of Deep Space missions anywhere in the Solar System would be a suitable task for the SKA, after its full implementation around 2020. We believe that this project will pave the way for future VLBI navigation observations of planetary and deep space missions, using the current VLBI network and new radio astronomical instruments under development.

## 6    CONCLUSIONS

A first assessment of the implication of the Square Kilometre Array for ESA space has been presented. It has been made clear that the SKA development should *not* be neglected with respect to the planning of future Deep Space missions. However all SKA concepts currently considered have limitations for use in Deep Space Telemetry together with Radio Astronomy. Further study is required to investigate all design elements of a SKA for the use in a DSN. A particular interesting area for DSN is the use of SKA event based; short specific operations like the reception of lander signals between launch and impact can be made possible with the use of the complete SKA sensitivity and resolution. A demonstration of the latter has been given with the assessment of the Huygens Probe with the use of existing, partially upgraded, telescopes and VLBI.

## 7    ACKNOWLEDGEMENTS